**Theory and experiment reveal unexpected calcium profiles in one-dimensional systems**

**S. L. Mironov**

Institute of Neuro- and Sensory Physiology, Georg-August-University, Göttingen 37073, Germany

*Short title:* Exponential and periodic profiles of diffusing calcium

*Corresponding author*: S. L. Mironov

Institute of Neuro- and Sensory Physiology, Georg-August-University, Göttingen 37073, Germany

Tel.: +49 (551) 39 54 72; E-mail: smirono@gwdg.de.

**Summary**

Calcium is an ubiquitous second messenger that triggers a plethora of key physiological responses. The events are initiated in micro- or nano-sized compartments and determined by the complex interactions with calcium-binding proteins and mechanisms of calcium clearance. Local calcium increases in the vicinity of single channels represent an essentially non-linear reaction-diffusion problem that have been analysed previously using various linearized approximations. I revisited the problem of stationary patterns that can be generated by the point calcium source in the presence of buffer and obtained new explicit solutions. Main results of the analysis of the calcium buffering are supplemented with pertinent derivations and discussion of respective mathematical problems in Appendices. I show that for small calcium influx the calcium gradients around established around channel lumen have quasi-exponential form. For bigger fluxes, when the buffer is saturated, the model predicts periodic patterns. The transition between the two regimes depend on the capacity of buffer and its mobility. Theoretical predictions were examined using a model one-dimensional system. For sufficiently big fluxes the oscillatory calcium patterns were observed. Theoretical and experimental results are discussed in terms of their possible physiological implications.

**Introduction**

Current view on the living cells consider them not as well mixed test tubes but rather specialized devices that consist of micro- or nanocompartments that are specialized on particular physiological responses and utilize complex biochemical cascades. Many fundamental responses are triggered by increases in calcium. It is an ubiquitous second messenger but its actions may be akin a double-edged sword. The key physiological processes such as synaptic transmission, secretion, contraction etc. are initiated by the fast, big and local calcium transients. Pathophysiological effects usually invoke global calcium increases when calcium loads last longer and cannot be sufficiently dampened by the calcium clearance mechanisms. Local calcium increases generated by the single calcium channels have big amplitudes and are well restricted. This feature may serve to activate the calcium binding proteins posited in the immediate vicinity of the channels. Theoretical justification of highly localized calcium transients has been set out by Neher (1983). He showed that the amplitude and width of local calcium transients are shaped by ubiquitous calcium-binding proteins or endogenously introduced calcium buffers. The concept of calcium micro- and nanodomains has been further used to explain the effects of calcium in the synaptic transmission and secretion (cf. Augustine et al. 2003; Falcke, 2004; Eggermann et al. 2011 for reviews). Application of imaging made such local calcium increases around single channels visible (Baylor et al. 2002; Wang et al. 2004; Beaumont et al. 2005; Shuai & Parker, 2005; Demuro & Parker, 2006; Tay et al. 2012, etc.). Although current optical resolution is not yet sufficient to precisely determine the temporal features and spatial patterns of single calcium transients, both theoretical and experimental vistas laid down a solid platform to understanding fundamental calcium-dependent events in molecular cell physiology.

Previous theoretical analyses correctly indicated a local character of calcium transients and stimulated new experiments. Unfortunately, in the sense of mathematical physics, they were not rigorous enough. Diffusion of calcium from the lumen of single channel into the cytoplasm is an essentially non-linear problem. It has been simplified using assumptions that calcium does not saturate cytoplasmic buffers, its binding is fast and irreversible. This allowed to reduce the reaction-diffusion (RD) problem to the linear ordinary differential equation (ODE) of the second order that describes the steady state calcium gradients around the inner mouth of the channel. The solution is simple and lucid, indicating exponential decay of calcium concentration from the channel lumen. A pitfall is that the levels of calcium around single channel are likely high enough to saturate the buffer. For example, for the single channel current of 1 pA the radial profile in the medium without buffer is $c(r) = A/r$ (Mironov, 1990; Stern, 1992). In the vicinity of channel lumen with radius $R = 1$ nm (Boda et al. 2007), the concentration of calcium is 1.6 mM. The estimate is close to 1.1 mM recently measured using sensor tethered to the calcium channel (Tay et al. 2012). Such calcium levels exceed the normal concentration of intrinsic calcium buffers in the cytoplasm (~0.2 mM) that should show significant saturation after channel opening. Thus bot theoretical and experimental estimates do not support the use of linearization in describing single calcium gradients. The point has been recognized previously and Stern (1992) indicated that calcium levels may not be controlled by the buffer in the immediate vicinity of the channels. Further corrections to linear treatment have been proposed (Stern, 1992; Pape et al. 1995; Naraghi & Neher, 1997; Smith et al. 2001; Falcke, 2004) but their applicability is limited to the calcium levels well below buffer saturation.

  To account for essentially non-linear problem of calcium buffering Wagner and Keizer (1991) introduced rapid buffer approximation (RBA) that allows to cast RD equations into single partial differential equation (PDE). The resulting equations are however non-linear and do not possess explicit solutions and need further approximating assumptions (Smith, 1996; Gin et al. 2006; Mironova and Mironov, 2007). Additionally, RBA has intrinsic limitations to describe the calcium gradients generated by single channels. The main problem that it assumes establishment of equilibrium between calcium and buffer. The off-rate constant of calcium unbinding from the most buffers is $k_{off}$ ~100 s$^{-1}$ and >10 ms is needed to achieve equilibrium. Calcium channels open normally for much shorter time and the equilibrium may never be reached. Therefore Stern (1992) and Naraghi and Neher (1997) concluded that calcium ions exiting single open channel likely confront pre-existed free buffer distribution. Calcium unbinding from the buffer occurs at much longer time-scale (>0.1 s) and the effects are important for the global calcium transients in the bulk of the cytoplasm.

I revisited the problem of steady state calcium profiles around single channels and solved non-linear equations without imposing restrictions on the magnitude of calcium influx or extent of buffer saturation. I present the main results in the text and theoretical derivations are detailed in Appendices where also other possible solutions of RD equations are analysed. For calcium fluxes below buffer capacity I obtained the two steady state solutions. The one describes 'exponential' calcium decay from the source and another one blows up at finite distances from the channel lumen. Analysis of the time-dependent problem indicated that only the former solution is likely stable and can be reached. When calcium flux exceeds buffer capacity calcium distributions demonstrate periodic patterns. Transitions between the decaying and oscillating patterns is defined by the concentration and mobility of the buffer. Theoretical predictions were tested experimentally in the model 1D-system. At small calcium fluxes calcium increases were localized. Increase in the calcium current produced periodic stationary profiles. The experimental data and theoretical analysis are discussed in the context of their possible biological implications.

**Results**

*Non-linear analysis of calcium profiles*

The problem of calcium distribution in the cytoplasm is common for chemical kinetics and can be casted into the set of nonlinear parabolic partial differential equations (PDE)

$$\partial C/\partial t \;=\; D\Delta C - \sum(k_{on,n}\, CF_n - k_{off,n}\, B_n)$$
$$\partial B_n/\partial t \;=\; d_n \Delta B_n + \sum(k_{on,n}\, CF_n - k_{off,n}\, B_n) \tag{1}$$
$$\partial F_n/\partial t \;=\; d_n \Delta F_n - \sum(k_{on,n}\, CF_n - k_{off,n}\, B_n)$$

that considers the interactions of multiple ($n$) cytoplasmic buffers with calcium. Here $\Delta$ stands for the Laplacian; $C$, $F_n$ and $B_n$ are the concentrations of calcium, free and calcium-bound forms of $n$th buffer; and $k_{on,n}$ and $k_{off,n}$ are the rate constants of calcium binding to and dissociation from the buffers. The ratio $k_{off,n}/k_{on,n} = K_{d,n}$ defines the dissociation constant of the buffer that indicates its affinity to calcium. $D$ and $d_n$'s are the diffusion coefficients of calcium and buffers, respectively.

I consider the radial diffusion from the single channel that corresponds to the boundary condition

$$dC/dr = -i/2\pi DFR^2 \quad \text{at } r = 0$$

where $R$ is the exit radius of the channel, $i$ is the single channel current and $F$ is the Faraday constant. Diffusion of calcium from the channel into the infinite cytoplasm has radial symmetry. It is convenient to make substitution $c = C/r$ that transforms the diffusion term $(1/r^2)\partial(r^2(\partial C/\partial r))$ into $\partial^2 c/\partial r^2$ and the boundary condition into

$$Rdc/dr - c = -i/2\pi DF \quad \text{at } x = 0 \qquad (2)$$

For $R = 1$ nm (Boda et al. 2009) the first term in (2) can be neglected and the boundary condition

$$c_o = i/2\pi DFR \qquad (2a)$$

then describes 1D-diffusion from the point source held at the constant calcium concentration $c_o = A$. In all derivations below I use 1D presentation with the independent space variable denoted as $x = r$. The formula for 3D-radial diffusion are then readily obtained by dividing 1D-solutions by the distance from the channel lumen.

Calcium buffering in the native cytoplasm can be well approximated by single buffer with apparent concentration $B_o \approx 0.2$ mM and dissociation constant $K_d \approx 0.4$ μM (Mironova & Mironov, 2008). In order to get a general insight into the problem I first consider the case when only one buffer is present and then extend the treatment by adding another buffer with slower mobility. In the derivations I use a non-dimensional form of RD equations. All concentrations are normalized to the total buffer concentration $B_o$. I use also the non-dimensional time $t = t/\tau$ ($\tau = 1/B_o k_{on} = 50$ μs, $k_{on}$ set to $10^8$ M$^{-1}$s$^{-1}$) and space variables $x = x/r_o$ ($r_o = \sqrt{(D/B_o k_{on})} = 0.1$ μm, $D$ for calcium is set to 200 μm$^2$/s).

In the presence of single buffer the diffusion of calcium from the channel is described by the two PDEs

$$c_t = c_{xx} - c(1-b) + \gamma b \qquad (3)$$

$$b_t = db_{xx} + c(1-b) - \gamma b$$

where $x$ is 1D-coordinate, $c$ and $b$ are the normalized concentrations of free and bound calcium, $d$ is the diffusion coefficient of buffer relative to that of calcium ($d = d_1/D$) and $\gamma = k_{off}/k_{on}B_o = K_d/B_o = 0.002$ is the normalized dissociation constant.

*Multiple 1D-calcium profiles*

Addition of the two equations in (3) gives

$$(c+b)_t = (c + db)_{xx}$$

From the uniqueness of solution to Laplace equation, a condition

$$c + db = const = A \qquad (4)$$

must be fulfilled in the steady state. The constant is defined by setting $c_o = A$ at the channel lumen to the calcium influx (2a). Then the concentration of bound calcium is

$$b = A/d - c \qquad (4a)$$

Eq. (4) is equivalent to that used by Pape et al. (1995) and Naraghi and Neher (1997) who assumed that the total flux of calcium (in free and bound form) is $(c + db)_x = 0$. They then linearized the system (3) to obtain steady state profiles. The explicit solution of complete non-linear problem is derived below. Using (4a) the system is transformed into

$$c_t = c_{xx} - c(1 - A/d + c/d + \gamma c) + \gamma A/d \qquad (5)$$

I consider first the case when $\gamma = 0$ and $d = 1$ (the simplifications are not essential and will be removed later on). In the stationary case this gives the ODE

$$c_{xx} = c(1 - A + c) \qquad (5a)$$

It is convenient to further rewrite it as

$$s_{zz} \pm s - s^2 = 0 \quad (6)$$

where the variables are $s = c/|1 - A|$ and $z = x\sqrt{|1 - A|}$. The signs (+) and (−) in (6) describe the cases $A < 1$ and $A > 1$ i. e. when the calcium flux is smaller or bigger than the buffer capacity, respectively. The first integration gives

$$s_z^2 = s\,(2s/3 \pm 1)$$

and the second one results gives

$$\pm z = \log \frac{\sqrt{(2s/3 + 1)} - 1}{\sqrt{(2s/3 + 1)} + 1} - \log \frac{\sqrt{(2s_o/3 + 1)} - 1}{\sqrt{(2s_o/3 + 1)} + 1} \qquad A < 1 \quad (7a)$$

$$\pm z/2 = \arctan[\sqrt{(2s/3 - 1)}] - \arctan[\sqrt{(2s_o/3 - 1)}] \qquad A > 1 \quad (7b)$$

The last term in the right-hand side stands for the integration constant ($w$) that depends from $s_o = A/|1 - A|$. Inversion of (7) gives the normalized concentrations as

$$s = \frac{3/2}{\sinh^2[(\pm z + w)/2]} = \frac{3}{\cosh[\pm z + w] - 1} \qquad A < 1 \quad (8a)$$

$$s = \frac{3/2}{\sin^2[(\pm z + w)/2]} = \frac{3}{1 - \cos[\pm z + w]} \qquad A > 1 \quad (8b)$$

Direct substitution of (8) into (6) proves that the formulas satisfy the boundary condition $c_{x=0} = A$. Explicit spatial dependencies of calcium gradients are obtained by replacement of the normalized concentration and space variables as $c = s\,|1 - A|$ and $z = x/\sqrt{|1 - A|}$.

Each Eq. (8) defines the two steady-state spatial profiles that are determined by ± sign in the argument and mirror each other. For (−) sign the solution in (8a) is ~1/*cosh* and decays exponentially in the positive half-plane (Fig. 1A) that reminisces the exponential solution of the linear ODE $s'' = s(1 - A)$ or $c'' = c$. It is a special case considered by Neher (1983) when free buffer is in excess and/or calcium loads are small ($A << 1$). Because in this case the space constant is $\sqrt{|1 - A|} \approx 1$, the concentration profiles are $s = exp(-x)$ or $c = A exp(-r/r_o)$.

The solution (8a) with (+) sign peaks at $z = w$ in the right half-plane (Fig. 1B). Such behaviour is typical for the blow-up solutions that often appear in RD field (cf. Galaktionov & Svirshchevskii (2007) for exhaustive summary and extensive literature). Such phenomena have numerous implications in the combustion, heat and mass transfer processes. For calcium gradients they are derived using formal mathematics and their role in the generation of calcium increases within living cells is tentatively unclear.

It logical to assume that calcium gradients around the channel always have the same form and are proportionally scaled with the amplitude of calcium influx. The solutions obtained show that the calcium gradients change their width in dependence from the space scaling factor $\sqrt{|1 - A|}$. Such variations are particularly prominent when the calcium flux approaches buffer capacity at $A = 1$ (Fig. 1A). Furthermore, when the calcium flux crosses this critical point, the spatial patterns become periodic (Fig. 1C). This simply reflects the fact that hyperbolic functions in (8a) are replaced by trigonometric functions in (8b). Such calcium patterns have not yet been documented in biological systems and possible reasons for this are delineated in Discussion. I did observe the oscillating calcium distributions in the model 1D-system (see Fig. 3 below).

The issue of whether calcium increases develop into the steady state profiles prescribed by (8a) and (8b) is analysed in Appendix A. I found that the two solutions with (+) and (-) sign in Eq. (8) produce into the same stationary patterns (the exponential one for $A < 1$ and periodic one for $A < 1$ shown in Fig. 1A and 1C, respectively). Kinetic analysis thus seemingly excludes a diverging (blow-up) solution with (+) sign presented in Fig. 1B. The conclusion may not be ultimate, because the kinetic analysis was relied upon the use of particular time-dependent function. Rigorous examination of 'blow-up' solutions is difficult and requires non-standard approaches (Ferreira et al. 2003), and I leave the issue tentatively open.

Removal of the first simplifying assumption ($\gamma = 0$) gave solutions presented in Appendix B. Because $\gamma$ values normally are small ($\gamma = k_{off}/k_{on}B_o = K_d/B_o \approx 0.4$ μM/0.2 mM = 0.002 << 1), the effects of calcium unbinding from the buffer does not significantly modify calcium gradients in the steady state.

*When the buffer diffuses slower than calcium*

The second simplifying assumption in obtaining (8) was $d = 1$ that corresponds to equal diffusion coefficients for buffer and calcium. This is not crucial, because Eq. (4) is transformed into Eq. (6) after defining the space and concentration variables as $z = x\sqrt{|1 - A/d|}$ and $s = c/|d - A|$. The solutions are then still determined by Eq. (8) given appropriate modifications.

Fig. 2 shows how the changes in $d$ modify the amplitude and width of calcium gradients. The examples were generated for $A = 0.1B_o$ and $2B_o$ that bracketed the critical point at $A = d$. It is seen that unless the diffusion coefficients for calcium and buffer differ by no more than 3-fold, the steady calcium distributions change very little. Decrease in $d$ makes the 'exponential' profiles wider for $A < d$, because space scaling factor $\sqrt{(1 - A/d)}$ gets smaller (Fig. 2A). Mechanistically speaking, a slower moving buffer may be not redistributed fast enough to effectively capture calcium steadily coming out of the channel. The effect may play an important role in determining the width of calcium nanodomains. The space scaling factor $\sqrt{(A/d - 1)}$ for $A > d$ increases with decreasing $d$ that sharpens the peaks in periodic calcium patterns and shorten the distance between them (Fig. 2B).

The major effect of slower buffer mobility is a shift of the critical point that separates the 'exponential' and periodic solutions to smaller values of calcium influx. It is defined as $A = d$ and when $d = 0.1$, the critical value is $A = B_o/10 \approx 0.02$ mM. As indicated in Introduction, the single channels likely generate bigger calcium increases that should give rise to periodic solutions, not yet described in the living cells. In Discussion I argue that they may be missed, because in radial profiles the amplitude of the peaks (Fig. 1C, on the right) successively diminish and the effect is amplified due to broadening of peaks during imaging (Appendix G).

*Experimental 1D-calcium profiles*

To test the predictions of the theoretical model I designed simple microscopic set-up and examined one-dimensional calcium RD patterns. I pulled fine micropipettes, filled with the indicator dye Mag-Fura Red (30 μM) and placed them into the bath with 1 mM $Ca^{2+}$ (the background electrolyte contained 145 mM NaCl and 10 mM HEPES at pH 7.4). The pipettes had elongated conical tips with ~100 nm openings (measured by the electron microscopy) and 50 ± 5 MOhm resistance. Iontophoresis evoked calcium entry into the pipette and indicator exit out of it. The diffusion coefficients in an aqueous solution are $D_{Ca}$ = 600 μm$^2$/s and $D_{Mag\text{-}Fura\ Red}$ = 300 μm$^2$/s (Hrabetová et al. 2009). Diffusion in sharp conical tips is equivalent to 1D-diffusion with twice smaller diffusion coefficient (Kalinay & Percus, 2006). Thus in this experimental configuration calcium and Mag-Fura Red had about the same apparent diffusion coefficients that corresponds to the case rigorously treated above. The space constants for radial diffusion of indicator into the bath were $r_o = \sqrt{(D_{Mag\text{-}Fura\ Red}/[Ca]_o k_{on})}$ = 0.06 μm and for 1D-diffusion of calcium into pipette $r_i = \sqrt{(D'_{Ca}/[Mag\text{-}Fura\ Red]_o k_{on})}$ = 0.33 μm. The experimental configuration I used is reversed in comparison with that occurs in the cytoplasm where calcium exits from nm-wide pore into the infinite medium. Attempts to model the latter case were not successful. When 1 mM calcium was in the pipette, its insertion into the bath contained 30 μM Mag-Fura Red led to immediate formation of precipitates that plugged the electrode tips.

Micropipettes were positioned under 63x objective lens of an upright microscope (Axioscope 2, Zeiss). For imaging I used 488 nm light from SLM Diodenlaser (Soliton, Gilching, Germany) and captured subsequent frames by cooled CCD camera (BFI Optilas, Puchheim) using ANDOR software (500 x 500 pixels at 12 bit resolution, 0.1 s acquisition time). Fluorescence signals were analysed offline with MetaMorph software (Universal Imaging Corp., Downington,

PA, USA) and custom-made programs. Signal were also deconvoluted to improve the spatial resolution in *x-y* plane (Mironov & Symonchuk, 2006). Calcium values were calculated from one-wavelength measurements using the standard approach (Grynkiewicz et al. 1985). The minimal and maximal fluorescence levels were obtained experimentally and the dissociation constant for calcium binding to Mag-Fura Red was set to 17 µM.

Mag-Fura Red in the calcium-free form had very weak fluorescence and the pipettes were barely distinguished from the background. Application of iontophoretic current increased the fluorescence around the tip and within the pipette. This indicated the binding of ejected Mag-Fura Red by calcium in the bath and calcium entry into the pipette. Increase in current produced several equidistant spots in the pipette tip (Fig. 3). Formation of stationary profiles had complex kinetics as shown by the kymographs in Fig. 3A. The spots lose and gain the intensity until the steady periodic patterns were established. Such behaviour has been described for so-called replicating spots in various theoretical and experimental RD models (Koch & Meinhardt, 1994; Vanag & Epstein, 2007). After switching off the current, the periodic steady patterns dissipated. The positions of peaks are well approximated by the maxima predicted by Eq. (8b). For bigger iontophoretic currents they were placed more densely (Fig. 3B). Measured distances between the spots had inverse square root dependence from the amplitude of calcium influx (Fig. 3C), also in accord with the theoretical model.

**Discussion**

Calcium acting as an ubiquitous second messenger initiates a plethora of key physiological reactions (Berridge, 1998). They are triggered by calcium binding to the specific proteins that stimulate particular signalling cascades. The targets involved in calcium-mediated synaptic transmission and secretion have low affinity calcium binding sites. This may help to dampen their spontaneous activation and assure reliable generation of synaptic events only by large amplitude (> 10 µM) calcium transients. Low affinity calcium signalling might have caused another problem.

When appropriately big cytoplasmic calcium increases occur in the bulk, diverse pathological processes e. g. programmed cell death (apoptosis) could be initiated. This may be a reason why physiologically relevant calcium-dependent responses are strictly compartmentalized and high calcium levels are achieved only in the immediate vicinity of its targets. The feature has been first recognized by Neher (1983) and since then the concept of calcium micro- and nano-domains received full acceptance and firm experimental confirmation (Baylor et al. 2002; Beaumont et al. 2005; Shuai & Parker, 2005; Demuro & Parker, 2006; Laezza & Dingledine, 2011; Shkryl et al. 2012; Tay et al. 2012). Nowadays calcium gradients produced by single channels are implicated in various physiological responses (Augustine et al. 2003; Cheng & Lederer, 2008; Eggerman et al. 2011).

Cytoplasmic buffers shape calcium distribution and such interactions are described by the reaction-diffusion systems. Many of studies in the field delivered important insights in understanding spatio-temporal behaviour of ecological and biological systems (Gray & Scott, 1990; Murray 2002; Petrovskii & Li, 2006; Grzybowski, 2009). They are yet not fully exploited for describing molecular processes in the living cells including calcium-mediated responses.

Calcium RD processes are characterized by the intrinsic time and space constants whose values are about 50 μs and 0.1 μm, respectively (see the first section of Results). After opening of single channels the steady calcium increases are established within <1 ms within ~1 μm from channel lumen that is confirmed by the optical tracking of single channel activities (Demuro & Parker, 2006). Neher (1983) predicted the exponential form of calcium gradients around the channels. He used the assumptions that after calcium exit out of the channel (i) it is irreversibly captured by the buffers and (ii) does not saturate them. This neatly allowed to put RD equations into single linear ODE of the second order and obtain exponentially decaying calcium gradients.

In most relevant cases the first assumption is valid, because calcium unbinding from the buffer is slow, takes >10 ms and can hence be neglected (Stern, 1992; Naraghi & Neher, 1997). I revisited the problem in Appendix B and came to a similar conclusion (the side result was derivation of new RD solutions in terms of Jacobi functions). The second assumption about the first-order kinetics of calcium binding has limitations that are important. As indicated in Introduction, calcium increase at the site of calcium exit into the cytoplasm is ~ 1 mM that is comparable or may even exceed the normal levels of cytoplasmic buffers (~ 0.2 mM). This inherent inconsistency in theory was recognized previously and various non-linear generalizations of calcium RD problem have been proposed (Wagner & Keizer, 1991; Stern, 1992; Smith, 1996; Smith et al. 2001; Tsai & Sneyd, 2007; Mironova & Mironov, 2007; etc.). In some of these studies the rapid buffer approximation (RBA) was used that conveniently cast RD system into single PDE (or ODE) of the second order. The use of RBA has been correctly criticized, because the equilibrium between calcium and buffer is unlikely to be reached during short openings of the channels (Stern, 1992; Naraghi & Neher, 1997). Non-linear character of calcium buffering was also treated using linearized formulations (Stern, 1992; Pape et al. 1995; Naraghi & Neher, 1997; Smith et al. 2001), but they also have limited applicability. The approximations are valid for calcium levels well below the dissociation constant of the buffer ($K_d \approx 0.4$ μM) or capacity ($B_o \approx 0.2$ mM), and actual calcium increases near the channel lumen are likely much bigger. Furthermore, I show in Appendix D that formal derivation of successive corrections to linear ODE solution produces series that poorly converge. Their truncation produces negative calcium levels even at moderate calcium fluxes.

Another approach is to simulate calcium transients (cf. Matveev et al. 2004) that has unlimited possibilities and allow to treat multiple buffers with detailed kinetics of calcium binding. Numerical experiments however pursue the partial differential equations (PDE) to converge to a particular solution. Diffusion equations are known to possess multiple solutions (Carslaw & Jaeger, 1957; Polyanin & Zajtsev, 2004) and they cannot all be generated in simulations. Integration algorithms dampen or reject some possibilities forcing to converge to a single solution. This point has been illustrated by Petrovsky and Li (2007) who discussed the examples of numerical and exact analytical treatments of the same problem of mathematical physics that gave distinctly different results.

I obtained exact solutions of the non-linear problem of steady state calcium distributions in the case of single buffer. They were derived using condition of uniqueness of solution to Laplace equation (4). It is equivalent to that used by Pape et al. (1995) and Naraghi and Neher (1997) who required zero total flux of free and bound calcium. Integration of Eq. (5) gave multiple steady state solutions enlisted in Eq. (8). When calcium influx is smaller than buffer capacity, one solution decays quasi exponentially (Fig. 1A) and another one diverges and blows up at finite distances from the channel (Fig. 1B). The kinetic analysis (Appendix A) indicated that only the former stationary solution can be attained, but a conclusion about impossibility of blow-up solutions may be not ultimate. They are well-known in various fields of physics (cf. Galaktionov & Svirshchevskii (2007), for recent survey), but require special tools for analysis of their spatio-temporal patterns (Ferreira et al. 2003).

When calcium flux is greater than buffer capacity, the steady state distribution attains periodic patterns with regularly spaced peaks. Their amplitudes are all equal in 1D case and decrease hyperbolically in the case of radial diffusion. For the theoretical analysis 1D- and 3D-radial problems are equivalent but the former are better suited for experimental studies. Using 1D-system I examined calcium binding to the fluorescent sensor Mag-Fura Red in model 1D system. Application of iontophoretic current drove calcium into the micropipette and produced periodic calcium patterns (Fig. 3A). They could be well resolved because the use of low buffer concentration increased the space constant of calcium buffering and well discernible calcium peaks were observed. Mag-Fura Red diffusing out of the pipette was saturated with calcium that produced fluorescence spot with sharp edges (Fig. 3B). In this case the space constant for calcium-buffer interaction was too small to reveal any spatial features.

Periodic spatial distributions are well known in physics and chemistry as Liesegang patterns (Liesegang, 1896; Jahnke & Kantelhardt, 2008). The estimates of calcium levels in the immediate vicinity of channels (~1.6 mM at the channel lumen, Introduction) are bigger than buffer capacity and should induce periodic calcium profiles. They have not yet been observed in the living cells that may have both theoretical and experimental reasons. First, under normal physiological conditions the distance between concentric shells and their a half-widths (FWHM) in periodic patterns are $2\pi r_o$ = 0.6 μm and 50 nm, respectively. This ideal case can be significantly distorted by imaging. I simulated the effects in Appendix G by convolving theoretical predictions with the Gaussian point-spread function (psf). For small calcium fluxes ($A < 1$) calcium gradients decrease their amplitude and broaden, and for $A > 1$ the secondary peaks in periodic patterns virtually disappeared. This indicates intrinsic experimental difficulties in observing periodic radial calcium patterns. Many measurements (cf. Beaumont et al. 2005; Demuro & Parker, 2006; Laezza & Dingledine, 2011) indicate extended calcium profiles where the main big peak is accompanied by a distinct shoulder at ~1 μm from the channel lumen. Such 'tails' may hide the secondary calcium peaks. Periodic patterns may have important physiological implications e. g. during calcium release from intracellular stores. It is mediated by single channels formed by ryanodine- and IP$_3$-receptors whose opening is facilitated by calcium (Cheng & Lederer, 2008). The channels often form clusters and secondary peaks from one channel may significantly contribute to the activation of neighbouring channels. After opening of ryanodine receptors single calcium gradients have also wide shoulders (Baylor et al. 2002; Shkryl et al. 2012) that may underline secondary peaks.

The analytical treatment indicated a crucial role of buffer mobility in shaping calcium transients and defining their pattern ('exponential' vs. periodic). When the diffusion coefficient of buffer decreases, the space normalization constant $1/\sqrt{|1 - A/d|}$ gets smaller that broadens calcium gradients (Fig. 2). The effect must be taken into account to correctly predict the width of calcium nanodomains. More critical is a shift of the critical point that separates the exponential and oscillating regimes. For equal diffusion coefficients of calcium and buffer it occurs when calcium flux is equal to buffer capacity ($A = 1$) and shifts to smaller calcium fluxes ($A = d$) in relation to the ratio of buffer and calcium diffusion coefficients (d ≤ 1).

Introduction of additional buffer complicates the analysis. One possibility is to cast multiple buffers into a single one that mimics the total buffer capacity of the cytoplasm (Mironova & Mironov, 2007). Another possibility is to take into account the fact that the on-rate constants of calcium binding are about the same[§]. The main obstacle in solving the problem of steady calcium gradient in the case of multiple buffers are their different diffusion coefficients. I considered the effects of slowly moving buffer on the distribution of calcium established in the presence of fast buffer. This case is physiologically relevant at it mimics frequently encountered situation when e. g. synthetic calcium indicator dye and intrinsic calcium buffer are both present in the cytoplasm. The diffusion coefficient of the former is close to that of calcium (Hrabetová et al. 2009) whereas for the latter it is by the order of magnitude smaller (Sanabria et al. 2008). Approximate solution of the problem showed that the shape of steady state calcium transients was dominated by the fast buffer and with almost negligible contribution of slowly moving buffer (Fig. SD).

In other words, consideration of the two differently moving buffers tempts to speculate that calcium gradients measured using synthetic calcium indicator dyes represents mostly their actions and the effects of intrinsic calcium binding proteins may be severely dampened. The time-course and spatial patterns of calcium changes recorded may be significantly distorted in comparison with those occurring in the intact cytoplasm when only native calcium binding proteins are present. The disturbances can be minimized using the dyes at lower concentrations, at best down to 10 μM. A better option is to apply genetically encoded calcium-binding proteins (Palmer & Tsien, 2006; Ohkura et al. 2012). First, they have the enhanced fluorescent yield, but much more important is that their mobility is close to that of intrinsic calcium binding proteins that served to design genetically encoded calcium sensors. This should induce much smaller disturbances in cytoplasmic calcium buffering and be better suited to depict the fine structure of single calcium gradients.

---

[§]Such notorious examples as EGTA and parvalbumin act as so-called 'slow calcium buffers', because they generate small quantity of the high-affinity calcium forms as argued by Pape et al. (1995) and Mironova and Mironov (2007).

It was previously noted (Stern, 1992; Pape et al. 1995; Naraghi & Neher, 1997) that immobile buffers should not modify steady state calcium profiles. I revisited the problem in Appendix F and derived the solution of the time-dependent problem. The calculations support previous notions and indicate that calcium gradients around the channel in the presence of fixed buffers must be much wider. The profiles in this case show strong time-dependence – they develop within a short time, spread out and saturate the immobile buffer.

A concept of calcium compartmentalization (Neher, 1983) gave rise to the concept of nano- and microdomains in calcium signalling (Augustine et al. 2003; Eggermann et al. 2011). It is now well supported by the measurements of single calcium transients (Baylor et al. 2002; Beaumont et al. 2005; Demuro & Parker, 2006; Laezza & Dingledine, 2011; Cheng & Lederer, 2008; Shkryl et al. 2012). In the last decade the optical resolution has been much improved, but the facilities may be yet limited to discern the fine structure of calcium gradients. Theoretical analysis indicates that calcium distribution around single channels may be more complicated than previously thought. Further enhancement of imaging resolution in combination with sensors tethered at different distances from the channel mouth (Tay et al. 2012) may reveal new features of calcium profiles around single channels in the native environment.

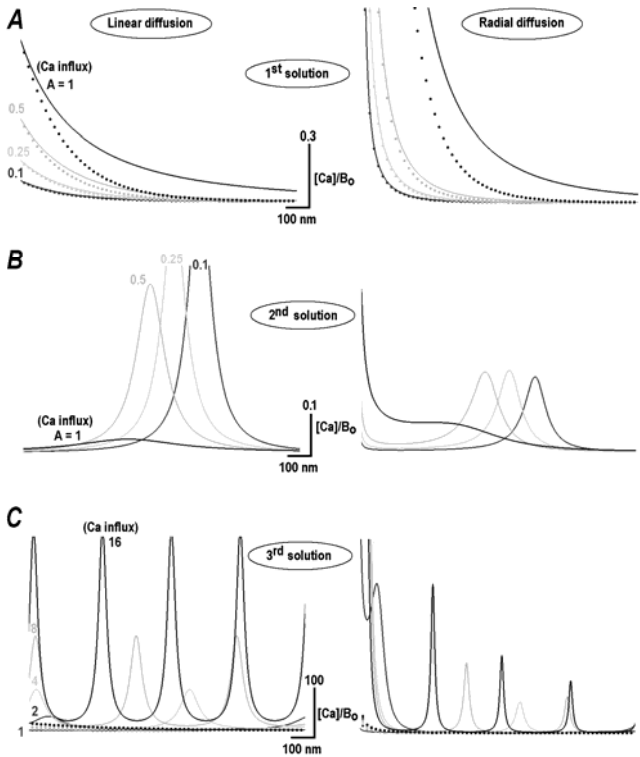

**Fig. 1. Theoretical steady state calcium profiles.**

Solid curves were calculated according to Eq. (8) and dotted curves plot the functions $A\exp(-r/r_o)$ (the space constant $r_o = 0.1$ μm), the solutions of linear ODE. *Left* - 1D-solutions for $A < 1$ (calcium influx smaller than buffer capacity). Panels (**A**) and (**B**) show the profiles for (+) and (-) sign in *sinh* argument in Eq. (8a), respectively. **C** – 1D-solution with (+) sign in *sin* argument in Eq. (8b) for $A > 1$ (calcium influx bigger than buffer capacity). *Right* – the radial solutions of calcium diffusion from the single channel into the cytoplasm. The curves were obtained from 1D-solutions that were divided by the distance from the inner mouth of the channel. Note different concentration scales in the panels and that calcium levels are presented normalized to the buffer capacity.

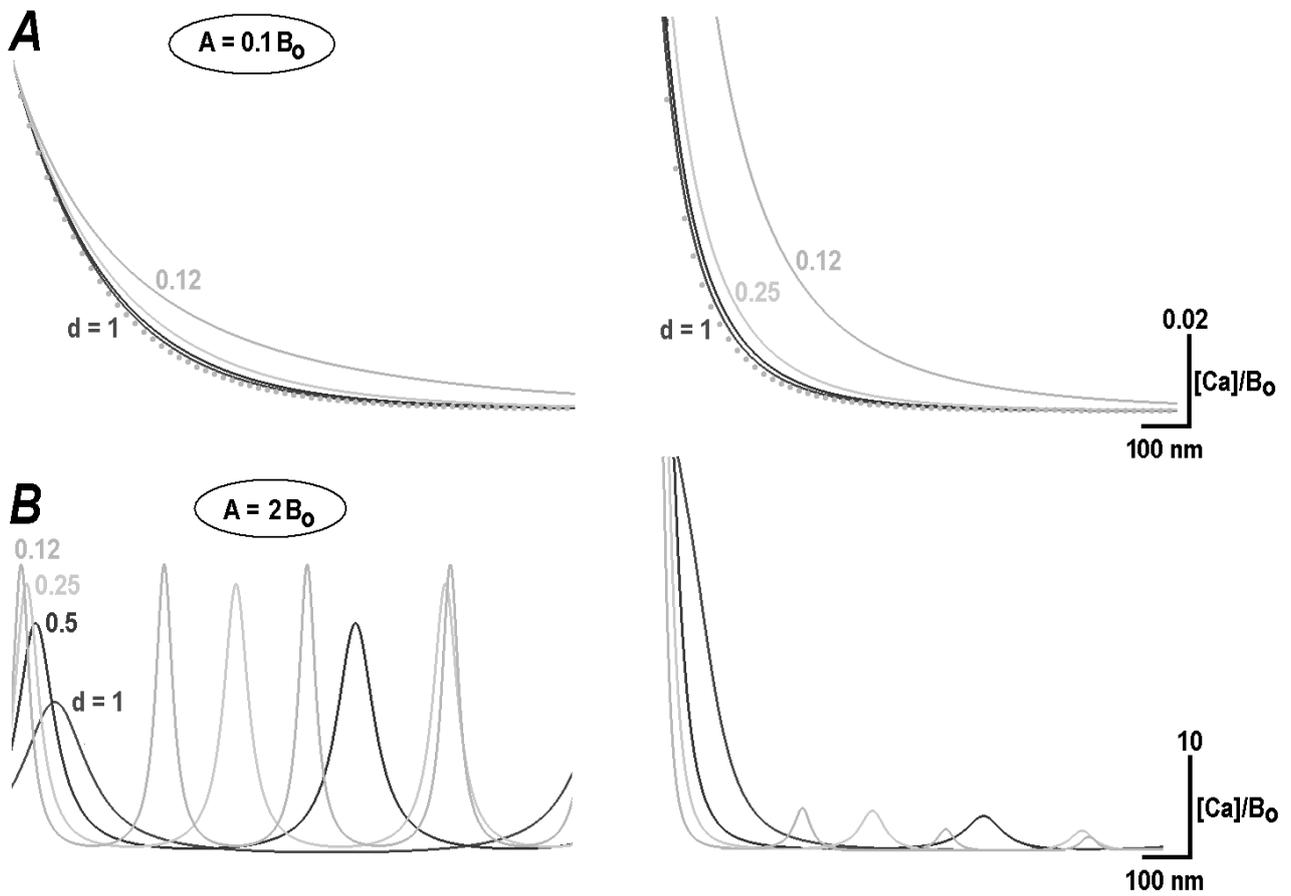

**Fig. 2. Steady state calcium profiles in the presence of the two buffers with different diffusion coefficients.**

1D- and 3D-radial solutions are shown in the left and right panels of *A* and *B,* respectively. The relative diffusion coefficients (*d*) of slowly moving buffer are indicated near respective curves. Calcium concentrations are normalized to the total concentration of buffer (≈0.2 mM).

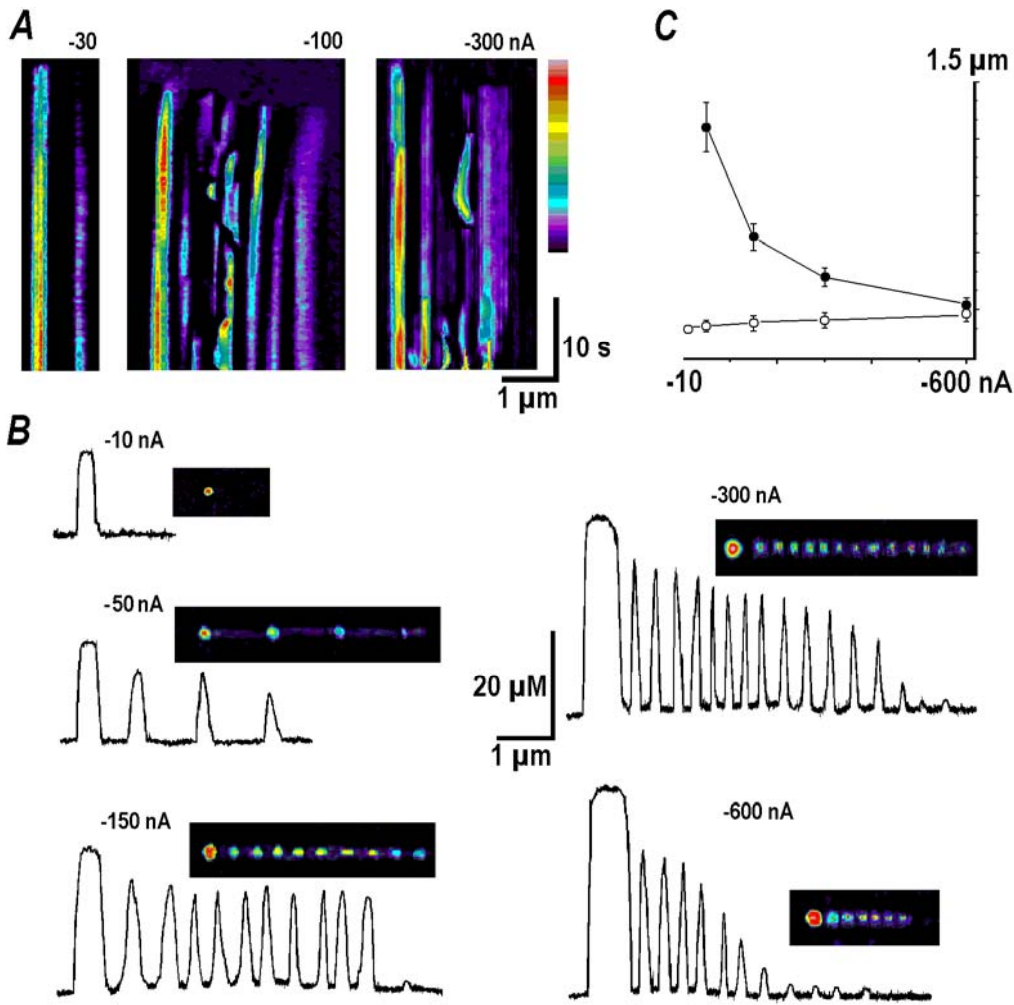

**Fig. 3. 1D-calcium profiles established by iontophoresis in micropipette tips.**

*A* – The kymographs show development of calcium increases during applications of steady iontophoresis currents whose amplitudes are indicated. Increases in Mag-Fura Red fluorescence are presented using pseudocolour coding. Note progressive appearance of several calcium spots in the interior of micropipettes that developed into the stationary calcium profiles.

*B* – Stationary calcium profiles. Fluorescence distributions are shown as images in the insets and as curves obtained from 1D-scans. Broad spots in the insets on the left and corresponding maxima in the scans indicate diffusion of the dye into the bath out of the pipette that overlapped with calcium increase in their tips. Subsequent spots and peaks show calcium increases within pipette interior.

*C* - Mean widths (±SD) of calcium spots in the bath (empty circles) and distances between successive calcium spots in the pipette (filled circles) plotted vs. iontophoresis current.

**Appendix A.** *Time-dependence of calcium gradients*

To analyse the kinetics of calcium profiles, I considered the time-dependent problem

$$s_t = s_x \pm s - s^2 \qquad (A1)$$

that corresponds to the steady state formulation in Eq. (6). The equation does not belong to Painleve type and is not integrable (Steeb et al. 1985). To find the solution I sought for appropriate probe functions that can converge to the steady state solution given by Eq. (8). Its form suggests several ways to introduce time-dependence and I found that the best one is

$$s = \frac{3f^2(t)}{ch - 1} \quad \text{for } A < 1$$

$$s = \frac{3f^2(t)}{1 - cs} \quad \text{for } A > 1 \qquad (A2)$$

where the notations

$$ch = cosh[\pm z\, f(t) + w] \qquad (A3)$$

$$cs = cos[\pm z\, f(t) + w]$$

Note that the arguments of cosines also contain the time-dependent function $f(t)$. When $f(t)$ increases from zero to 1, calcium increases near the point source first, then spreads out and attains the patterns predicted by Eq. (8). After insertion of (A3) into (A2) I obtained the first order ODEs

$$f_t = \frac{f^2(f^2 - 1)(ch + 2)}{2f(ch - 1) - 3f^2 z\, sh} \quad \text{for } A < 1$$

$$f_t = \frac{f^2(f^2 - 1)(cs + 2)}{2f(cs - 1) - 3f^2 z\, ss} \quad \text{for } A > 1 \qquad (A4)$$

where $sh = sinh[\pm z\, f(t) + w]$ and $ss = sin[\pm z\, f(t) + w]$. (A4) can be integrated that is not very useful, because this gives implicit expressions of the type $t = F(f(t), x)$ contained hyperbolic and trigonometric *sine* and *cosine* integrals. Such functional expressions cannot be inverted to obtain explicit time-dependences. I integrated (A4) numerically and found that $f(t)$ with (+) sign changed it to (-) and then converged to the stationary solutions given by Eq. (8) shown in Fig. 1A and 1C for $A < 1$ and $A > 1$, respectively.

I also numerically integrated the original PDE (A1). For $A < 1$ the calcium distributions converged fast to the steady state, but in the case of $A > 1$ the convergence was not monotonous. The amplitudes of peaks in the periodic patterns changed with time until the steady state was established. Such behaviour was also observed in the experiments (Fig. 3A) and the effects may be related to the replication spots observed in some physical models and simulations (Koch & Meinhardt, 1994; Vanag & Epstein, 2007).

Kinetic analysis thus does not support the existence of the 'blow-up' solutions (Fig. 1B), but does not exclude them completely. They may appear when other time-dependent probe functions are used. The analysis of 'blow-up' behaviour is notoriously difficult, because the standard algorithms of PDE integration dampen such solutions and special approaches have to be used for their examination (Ferreira et al. 2003).

**Appendix B.** *Reversible calcium binding*

To analyse the effects of calcium unbinding on the steady state profiles I presented Eq. (3) as

$$s'' - (s+1)s + \gamma_1 = 0, \qquad \gamma_1 = \gamma A/(1 - A + \gamma)^2 \quad \text{for } A < 1$$

$$s'' + (s-1)s - \gamma_2 = 0, \qquad \gamma_2 = \gamma A/(A - 1 - \gamma)^2 \quad \text{for } A > 1 \qquad (B1)$$

where the normalized concentration and space variables are now $s = c/|1 - A + \gamma|$ and $z = x\sqrt{|1 - A + \gamma|}$. Eqs. (B1) differ from Eqs. (6) by the constant term and their general solutions are given by the double-periodic Weierstrass functions (Abramowitz & Stegun, 1972; Atlas of Functions, 2009). Such presentation is not very useful, because no convenient approximation of Weierstrass functions exists. I sought for solutions of (B1) using related Jacobi functions. They are specified by the index

*m* and for *m = 1* they are simply hyperbolic functions. For example, a correlate of *1/sinh* in (8a) is $ds(z|m)$. Insertion of the probe function

$$s = b + a \cdot ds(\alpha z | m)$$

into (B1) defines $a = (2m – 1)/(5m^2 - 5m^2 + 2)$, $b = A(2m – 1)/3$, and $\alpha = \sqrt{(A/3)}$. For *m = 1*, the parameter values are $a = 3\alpha^2$, $b = (\alpha^2 - 1)/2$ and $\alpha^2 = \sqrt{(1+4\gamma)}$. Using them the normalized concentrations the solutions can be presented as

$$s = \frac{\gamma}{2} + \frac{3(1 + \gamma)^2}{\cosh[\pm\alpha z + w] - 1} \qquad A < 1$$

$$s = \frac{\gamma}{2} + \frac{3(1 + \gamma)^2}{\cos[\pm\alpha z + w] + 1} \qquad A > 1 \qquad (B2)$$

For $\gamma \to 0$ the expressions transform into Eqs. (8). Of note, Jacobi functions with *m > 1* also satisfy (B1). At present it is difficult to assess the significance of such solutions. I simply consider this option as another manifestation of the fact that diffusion equations possess multiple solutions (Carslaw & Jaeger, 1957).

The results indicate that the main effects of calcium unbinding are the increase in the amplitude and width of steady state calcium transients as well as the appearance of small offset. Because in most cases *γ* values are very small (*γ* is defined as $k_{off}/k_{on}B_o = K_d/B_o \approx 0.4$ μM/0.2 mM = 0.002 << 1), the effects of calcium unbinding from the buffer should have little influence on the stationary calcium distributions.

**Appendix C.** *A case of two buffers*

Difficulties arise when multiple buffers are present and straightforward integration of RD equations is no more possible. Because the on-rate constants of calcium binding to the buffers are close to the diffusion limit and can be set equal, the general system in Eq. (1) can be normalized in the same way as in obtaining Eq. (3). The steady state calcium profile is determined by the set of (*n* + 1) equations

$$0 = c_{xx} - c(1 - \sum b_n) + \gamma_n b_n \tag{C1}$$

$$0 = d_n b_{nxx} + c(1 - b_n) - \gamma_n b_n \ldots$$

and the condition of uniqueness of solution to Laplace equation is now

$$c + \sum d_n b_n = A \tag{C2}$$

The problem is that the diffusion coefficients of buffers are different and the individual equations in (C1) cannot be uncoupled. Fortunately, in most relevant physiological cases the problem can be reduced to consideration of only two buffers with different mobilities. To the first class belong the synthetic calcium indicator dyes (Mag-Fura Red, fluo-4 etc.), chelators (EGTA, BAPTA) and small molecules (ATP, ADP) that all have similar diffusion coefficients close to that of calcium with maximally two-fold difference. Another class of buffers is represented by the calcium binding proteins and genetically encoded sensors. They are bulky proteins and have ~10-fold lower mobility. I pooled the species within the two binding classes together and consider a three-component system - calcium and two buffers with different diffusion coefficients $d$ and $D$ at concentrations $b$ and $B$, respectively. For the analysis I used singular perturbation approach.

The condition (C2) can be written as

$$c + db + \mu DB = A \tag{C3}$$

Because $D < d$, I introduced in the last term a small parameter $\mu < 1$ and used it further to expand the concentrations as

$$x = x + \mu x_1 + \mu^2 x_2 + \ldots \tag{C4}$$

where $x = c, b,$ or $B$. It appeared that already consideration of the first order correction is sufficient to drawing quite general conclusions. Insertion of (C4) into the equation for slowly moving buffer in system (C1) gives

$$\mu D(B + B_1)_{xx} = (c + \mu c_1)(T - B - \mu B_1) - \Gamma(B + \mu B_1) \tag{C5}$$

where $\Gamma \ll 1$ is the 'dissociation' constant and $T$ is the fraction of the slowly moving buffer (all concentrations are normalized to the total buffer capacity). In zero order by $\mu$ (C5) gives

$$0 = c(T - B) - \gamma B \tag{C6}$$

that simply describes the equilibrium between calcium and slow buffer. Then in zero order the concentration of calcium bound by the slow buffer is $B = cT/(c + \Gamma)$. For normal $[Ca]_o \approx 0.1$ µM, $K_d \approx 0.4$ µM and $TB_o \approx 0.2$ mM. This means that about 20 % of slow buffer is calcium bound. In zero order by $\mu$ the first Eq. (C1) is

$$0 = c_{xx} - c(1 - b - B) + \gamma b + \Gamma B \qquad (C7)$$

When the effects of calcium unbinding during calcium influx is neglected (see Appendix B), this ODE for calcium is equivalent to Eq. (5a) where the calcium influx ($A$) is reduced by the total concentration of free slow buffer. This may be considered as another manifestation of the notion (Stern, 1992) that immobile (or fixed) calcium buffers do not influence the form of stationary profiles and only scale their amplitude. Calcium profiles in zero order are thus given by Eqs. (8) after appropriate correction for the actual $A$.

Diffusion coefficient of the slowly moving buffer appears only in the first order by $\mu$. Using zero order concentrations $c$ and $b$ (with $c + db = A$) and the first order condition $c_1 + db_1 + DB = 0$ in (C3), I express $b_1$ through $c_1$ and $B$ and obtain inhomogeneous ODE of the second order

$$s_{1xx} = s_1(1 + 2s_1) + f(s_o) \qquad (C8)$$

Here again the normalized variables are used and other notations are $f(s_o) = \alpha s_o^2/(s_o + \beta)$, $\alpha = DT/(dt - A)$, $\beta = \gamma D(t - A/d)$ where $t$ is the fraction of the fast moving buffer. The solution of (C8) can be written as

$$s_1 = GS - S \int (dz/S^2) \int S f(s_o) \, dz \qquad (C9)$$

(Kamke, 1969), where $S = ch/sh^3$ is the solution of homogeneous ODE ($S$). Other notations are $ch = cosh(y)$ and $cs = cos(y)$ and the argument $y = (\pm z + w)/2$, see Eqs. (8). The denominator in $f(s)$ contains $\beta \sim \gamma D$ that is very small in comparison with $s_o$ and the free term can be approximated as $f(s_o) \sim \alpha s_o = 3\alpha/2sh^2$. After integration of (C8) the first order correction can be written as

$$s_1 = (3\alpha/8)[(1/sh^2) + (ch/sh^3)(y_o - y - th_o)] \qquad (C10)$$

The constants $y_o = w/2$ and $th_o = tanh(y_o)$ are determined to satisfy the condition $s_1 = 0$ at $x = 0$ that defines the constant $G$ in (C9).

Fig. SC shows the effects of slowly moving buffer on calcium profiles. Calculations were made for calcium fluxes below buffer capacity ($A < 1$) and $D = 0.1$ that can imitate physiologically relevant cases. It is seen that the effects of the slowly moving buffer in the first order ($s_1$) are small and become evident only for relatively big calcium fluxes when $A > 0.1$.

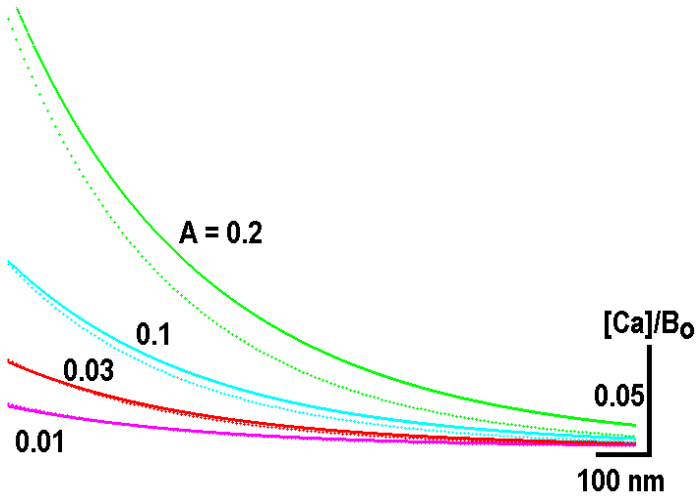

**Fig. SC. Steady state calcium profiles in the presence of two buffers.**

Solid curves were obtained in the presence of only fast moving buffer and the dotted curves were calculated after addition of the slowly moving buffer at the same concentration. Calculations were made using Eq. (C10) with $D = 0.1$. The curves show calcium profiles for different values of the normalized calcium influx ($A$) indicated near respective curves.

**Appendix D.** *Problems of linearized approximations*

This section is only to demonstrate some inconsistencies that arise when the problem of calcium buffering is solved after reduction to the linear ODE. It predicts the exponential profile of calcium around the channel in the form $c_o = Ae^{-x}$. Integration of equation for calcium-bound buffer $db_{xx} = -c$ defines then its profile as $b_o = (A/d)(1 - e^{-x})$. Therefore a proper ODE to describe calcium distribution should read as

$$c_{xx} = c(1 - b) = Ae^{-x}(1 - A/d - Ae^{-x}/d) = A(1 - A/d)e^{-x} + A^2 e^{-2x}/d \qquad (D1)$$

Its integration gives

$$c = A(1 - A/d)e^{-x} + (A^2/4d)e^{-2x} \tag{D1a}$$

The result shows that for $A \ll 1$ the concentration decays as $exp(-x)$ but when $A$ is close to $d$, the calcium decline is faster and becomes proportional to $exp(-2x)$. For $A > d$ calcium concentration may be negative.

Taking (D1) as zero order correction and using the identity $C + dB = 0$ to express $B = -C/d$, the ODE for the first order correction can be written as

$$C_{xx} = C(1 - b_o) - Bc_o = C(1 - A/d + 2Ae^{-x}/d) \tag{D2}$$

Using again normalized concentration and space variables $S = C/(1 - A/d)$ and $z = x\sqrt{(1 - A/d)}$ and introducing $\beta = 2A/(d - A)$, Eq. (D2) can be presented as

$$S_{zz} = S(1 + \beta e^{-x})$$

Its general solution is expressed via the Bessel functions

$$C = C_1 J_2(2i\sqrt{\beta}\, e^{-z/2}) + C_2 Y_2(2i\sqrt{\beta}\, e^{-z/2}) \tag{D3}$$

(Polyanin & Zaitsev, 1995). They have imaginary arguments whose real part $<1$. The behaviour of this solution can be qualitatively understood by considering the properties of the Bessel functions (Atlas of Functions, 2009). First, the second term ($Y_2$) diverges at small values of argument that corresponds to the large distances. In this sense it mimics the behaviour of the second solution of the linear ODE $c_{xx} = c$ that increases exponentially (blows up) and I leave it out of consideration. The first term is the modified Bessel function, $J_2(iy) = -I_2(y)$. At small values of argument ($e^{-x} < 1$) it is well approximated as

$$I_2(y) \approx (y^2/8)[1 + (y^2/24)]^4 \tag{D4}$$

Replacing $y^2$ with the argument of the Bessel function, the leading term of the first order correction is $-[A^2/d(d - A)]\, exp[-x\sqrt{(1 - A/d)}]$. For $A/d < 0.2$ this term is much smaller than that given by zero order correction (D1a), but for $A/d > 0.5$ it starts to dominate that makes the sum (D1a) + (D4) negative. For bigger fluxes, when $A > d$, the arguments of Bessel functions are real and they must show decaying oscillating behaviour akin Eq. (8b).

**Appendix E.** *Calcium profiles in the rapid buffer approximation (RBA)*

The assumption that calcium-buffer equilibrium is established instantaneously leads to RBA approximation that has been often applied for the analysis of calcium distribution in the cytoplasm (Wagner & Keizer, 1991; Smith 1996; Gin et al. 2006; Mironova & Mironov, 2007). In this section I consider some consequences of RBA that have not been described or attracted sufficient attention in the literature. RBA defines the concentration of calcium-bound buffer as $b = c/(\gamma + c)$ (the normalized form is used again). RD equation for calcium is then

$$c_t = c_{xx} - c/(\gamma + c)$$

or

$$s_{t'} = s_{x'x'} - s/(1 + s) \qquad (E1)$$

where $s = c/\gamma$, $t' = t/\gamma$ and $x' = x/\sqrt{\gamma}$. In the steady state direct integration of (E1) gives transcendental equation for concentration. The explicit solution cannot be obtained and numerical evaluation indicates that the solution decays about exponentially.

I attempted to solve (E1) using singular perturbation approach and expand

$$s = s + \mu s_1 + \mu^2 s_2 + \ldots \qquad (E2)$$

Insertion of (E2) into (E1) and collecting the terms with the same power by $\mu$ gives subsequent inhomogeneous ODEs whose solutions are

$s_1 = A \exp(-x')$

$s_2 = A^2[(1/3) \exp(-x')/3 - (1/3) \exp(-2x')]$

$s_3 = A^3[-(5/8) \exp(-x') - (1/9) \exp(-2x') + (1/24) \exp(-3x')]$

$s_4 = A^4[(11/18) \exp(-x') - (5/4) \exp(-2x') - (4/9) \exp(-3x') - (5/4) \exp(-4x')] \ldots \qquad (E3)$

The calculations show that even the sum of first two terms for $A < 1$ is not always positive and inclusion of further corrections makes the situation even worse. Thus application of perturbation approach is not very useful in RBA analysis.

When the reaction term in RBA is approximated as $s/(1 + s) \approx s(1 - s)$ (it is valid when the buffer is maximally half-saturated), Eq. (E1) can be written as

$$s_t = s_{x'x'} - s(1 - s) \tag{E4}$$

This is a seminal equation that has been derived by Fisher (1937) and Kolmogoroff-Petrovskii-Piscounoff (1937) to describe the expression of genes and spread of populations as the travelling waves. Many exact solutions of this and related equations were obtained as summarized by Polyanin and Zajtsev (2004). Surprisingly, calcium transients in the living cells have been never related to the F-KPP equation although it stems directly from RBA. Travelling wave solutions should describe the spread out of calcium from the plasma membrane that gradually fills the cytoplasm. Using parameters typical for the living cells (see the main text) I estimated the velocity of F-KPP solution of about 15 μm/s.

**Appendix F.** *Time-dependent solution in the case of immobile buffer*

When the buffer does not move, the equation for buffer in the steady state contains only reaction term $c(1 - b)$ that must be identically zero. This means that the ODE for calcium is $c_{xx} = 0$ that has only trivial constant solution. Mechanistically speaking, because the buffer cannot diffuse, calcium spreads out from the source until it fills the whole medium.

My interest to this problem was that in this special case a simple time-dependent solution can be obtained. Eq. (3) in the main text can be written as

$$c_t = c_{xx} - cf \tag{F1}$$
$$f_t = -cf$$

where the buffer diffusion coefficient $d = 0$ and $f = 1 - b$ is the normalized free buffer concentration. (F1) is solved using the approach proposed by Britton (1991). Introducing the variable $y = \int c\,dt$, I recast the system (F1) into single PDE

$$y_t = c_o + f_o[1 - e^{-y}] + y_{xx} \tag{F2}$$

where $f_o$ and $c_o$ are the initial concentrations of free buffer and calcium, respectively. It is seen that differentiation of (F2) produce exactly the first equation (F1) when the identities $c = y_t$ and $f = f_o e^{-y}$ are taken into account. Polyanin and Zajtsev (2004) give a particular solution of (F2) in the form of travelling waves. They have complex parameters and should correspond to the periodic solutions. Another type of solution is obtained by noting that, when calcium steadily accumulates, the exponential term $e^{-y}$ tends to zero. Then (F2) can be presented as

$$y_t = c_o + f_o + y_{xx} \tag{F3}$$

After it differentiation a simple diffusion equation

$$c_t = c_{xx} \tag{F3a}$$

is obtained. For the initial condition $c(x,0) = 0$ and boundary condition $c(0,t) = A$ it has the solution

$$c(x, t) = A\, erfc(x/2\sqrt{t}) \tag{F4}$$

The second equation $f_t/f = -c$ in (F1) defines free buffer concentration

$$f(x, t) = f_o\, exp(-y) = f_o\, exp(-\int c(x, t)dt) \tag{F5}$$

The time-dependencies of calcium and free buffer concentration are plotted in Fig. SF. It is seen that calcium increases first near the source, spreads out and saturates the buffer.

This approach may be used to analyse the case when the mobility of buffer is non-zero but very small. (F1) can be written as

$$c_t = c_{xx} - cf \tag{F1}$$

$$f_t = \mu d f_{xx} - cf$$

with $\mu d \ll 1$. This formulation might be a starting point for application of singular perturbation technique. Unfortunately I found that even for the first order correction one gets a system of two non-linear ODEs that is not generally solvable.

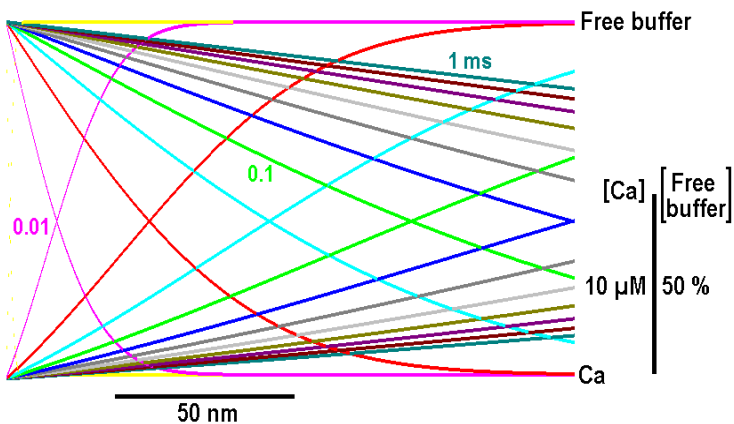

**Fig. SF. Concentration changes in calcium and free immobile buffer.**

The traces were calculated using Eq. (F4) for calcium 'influx' = 20 μM at progressively increased times (some are indicated near correspondingly coloured curves). Note that calcium gradually increases and saturates the buffer.

**Appendix G.** *Blurring single calcium gradients during imaging*

To estimate possible effects of optical resolution on single calcium gradients around the channels I convolved the radial calcium profiles given by Eqs. (8) with the Gaussian point spread function *psf* = $(\alpha/\sqrt{\pi})exp(-\alpha^2 x^2)$. Fig. SG shows that the main effects were a decrease in the amplitude and broadening of calcium gradients. This occurred both for 'exponential' and periodic patterns. The latter was most drastically changed such as initially well resolved peaks virtually disappeared. This was observed already for psf's with full-width-half-maximum (FWHM) = 0.2 – 0.3 μm that is about the best optical resolution of currently available custom microscopes.

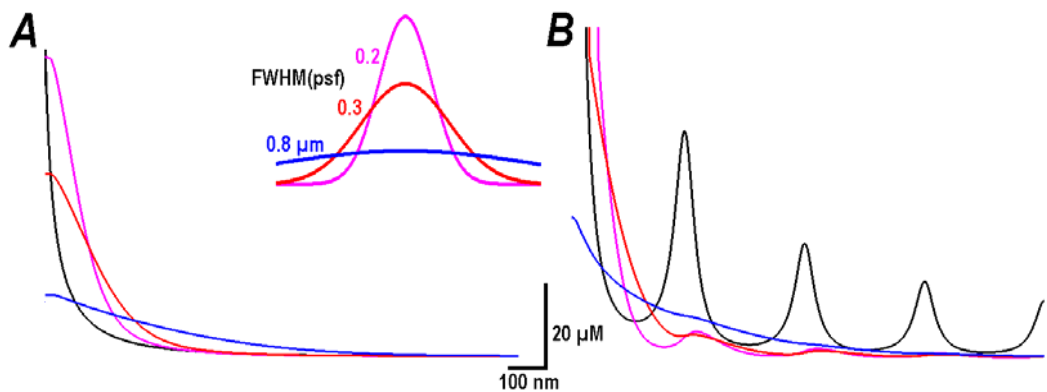

**Fig. SG. Smearing of theoretical calcium profiles.**

Black curves present theoretical 'exponential' (***A***) and periodic calcium profiles (***B***). Coloured traces show their transformation after convolution with the Gaussian psf's shown in the inset. Respective full-width-half-maximum (FWHM) values are as indicated.